# Interaction with tilting gestures in ubiquitous environments

Ayman Atia and Jiro Tanaka[1]

[1]Department of computer science, University of Tsukuba, Tsukuba, Japan
ayman@iplab.cs.tsukuba.ic.jp, jiro@cs.tsukuba.ac.jp

## Abstract

*In this paper, we introduce a tilting interface that controls direction based applications in ubiquitous environments. A tilt interface is useful for situations that require remote and quick interactions or that are executed in public spaces. We explored the proposed tilting interface with different application types and classified the tilting interaction techniques. Augmenting objects with sensors can potentially address the problem of the lack of intuitive and natural input devices in ubiquitous environments. We have conducted an experiment to test the usability of the proposed tilting interface to compare it with conventional input devices and hand gestures. The experiment results showed greater improvement of the tilt gestures in comparison with hand gestures in terms of speed, accuracy, and user satisfaction.*

## Keywords

*Interaction in ubiquitous environments, Tilt gestures, Human computer interaction*

## 1. Introduction

Human gestures are typical examples of nonverbal communication and help people to communicate smoothly. The substantial developments made in smart devices and small sensors are directly affecting the arena of human-computer interaction and smart ubiquitous environments. As a result, many researchers have considered using sensors to capture a user's environment and the human's intentions in order to get more intuitive interactions. A typical use of sensors in our daily life is to capture user activity [1] [2] or interaction with objects [3] [4].

Interaction in the ubiquitous environment could suffer from the absence of direct conventional interaction methods like keyboards, mice, and direct control devices. Conventional methods are designed mainly for stationary interaction situations. Using a camera to capture hand gestures depends on the lighting of an environment and requires one to face the camera, which limits mobility and involves privacy issues. Another method of interaction is using full hand gestures in ubiquitous environments. Some situations, like being on public transportation, however, are not conducive to full hand movements due to limited space. Hand gestures could also have unintended social implications based on the shape of the gesture [5] [6]. Finally, when users interact with applications using hand gestures for long periods of time, they may suffer from hand tremors and pain, to an increase in error rates.

The above issues relating to hand gestures for remote interactions have brought about the use of tilt gestures, as they depend merely on moving the wrist. People use human tilt hand gestures to express directions or to alter the position of one object in hand [7]. Many advantages exist for tilt gestures over hand gestures. First, they require less space to be preformed. Second, they can be easily memorized because they have a fixed number of positions. Third, the gestures can be performed at a faster speed in comparison with full hand gesture movements because they depend on moving only the wrist. Tilt gestures, however, also possess some research challenges for ubiquitous environment interaction, as they are limited in the number of actions they can perform. Furthermore, the interaction accuracy must be high, and the recognition engine must distinguish among different tilt directions and levels performed by a user. Finally, the suitable applications that benefit from tilt gestures in ubiquitous environments should be further studied. Users in some situations might be involved in other activities or may not have free hand. Consequently, we use everyday objects like pens, bottles, books, headphones, balls, cellular phones, and computers to act as input devices for the users.

## 2. RELATED WORK

Many techniques using hand gestures that are captured and tracked by a camera are discussed in the literature. Sawasaki et al. [8] and Azaz et al. [9] developed a camera-based system that captures the position of the hand and applies filtering techniques to convert it into mouse positions. Using a camera, however, could result in decreased accuracy and low processing power, and may be difficult to set up in ubiquitous environments. Another method to capture full hand gestures is through sensors. Lee et al. [10] developed the I-throw system to integrate location awareness and hand gestures. Farella et al. [11] developed a glove using bending sensors for hand interaction. Accelerometers have been used by Kela et al. [12] to capture full hand gestures and thereby help users interact with designing application. They considered implementing full hand gestures with limited commands that help designers interact with 3D modelling applications. Wobbrock [13] designed a $1 recognition engine for recognizing and classifying hand gestures using cheap devices.

Tilting operations for small screen interfaces were introduced by Rekimoto [14], who showed the advantages of having a tilt interface as an input device, especially in field work with the use of one-hand interaction. Furthermore, they pointed out the advantages of the tilt interface in the interaction with small screen devices that are not compatible with a pen device. However, the developed prototype system depends on combining tilt and normal device buttons for interaction. The lack of any input device present a challenge in the ubiquitous environment that requires further research. Levin et al. [15] reported a study on embedded accelerometers as a gesture interface for an extremely small computer using tilt and shake gestures for drawing. Lee et al. [16] proposed a tilting table as a movable screen to interact with several applications such as "Beadball" and "Cross-Being: Todd" tables, tilt maps, and bio browsers. They used the tilt gestures to move a display screen in four directions only. Hence, we thought about the suitable number of tilt gestures to be executed per application. Cho et al. [17] explored the use of the tilting gesture for a multi-context photo browsing on mobile devices and addressed some issues of the tilt interface, such as overshooting or a partial image problem.

Partidge et al. [18] proposed a tilt device using accelerometer-supported text entry for very small devices. They proposed a wristwatch with buttons for selecting the letter pad and use of the tilt to choose among letters. Their device, however, requires using both hands to control the device. Jani et al. [19] proposed a scan-and-tilt interaction technique for more natural interaction in museum guide applications. Andrew et al. [20] studied the variability in wrist-tilt accelerometer-based gesture interaction; they showed an increase in variability of motions upwards from the centre in comparison to downwards. They analyzed trajectories of accelerometer values and put in place design standards for tilt-based hand interactions. During their experiments, they controlled mouse cursor movements to hit targets using tilt directions. Antii et al. [21] proposed a controller for an "on the move" mobile media player using gestures and audio metaphors. They showed the importance of using the gestures to control the medial player while users are involved in other activities, such as walking or running. They pointed out the difficulty of operating small buttons inside small devices and the need for visual focus to control the media player buttons. They showed a limited number of gestures to command sound-based basic menus. However, they did not propose a method to control a menu with more functions and commands.

To the best of our knowledge, the number of targets and the applications suitable for tilt gestures in ubiquitous environment have not been studied. In this study we explored applications and mapping techniques for tilt gestures. We also studied the suitable range of commands that can be achieved by tilt gestures.

## 2. SYSTEM OVERVIEW

In our research, we used a coin sized small sensor, as it can be attached easily to various objects. Figure 1 shows the 3D accelerometer sensor [22] and its corresponding axes. It has a built-in 3D accelerometer, angular rotation rate sensor, and temperature sensor. The sensor is 20 g in weight 39 (w) x 44 (h) and 12 mm (d) in size. It has Bluetooth connectivity and battery power of up to 4 hours.

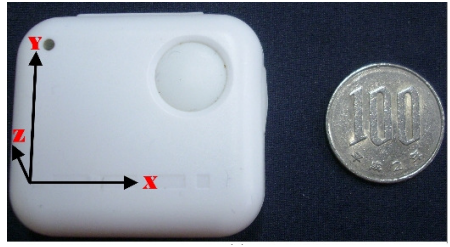

Figure 1. Coin sized 3D accelerometer sensor and its corresponding axes.

The sensor sends data as a pattern of signals that represent the acceleration measurements applied over the sensor's three axes. When a user tilts the 3D accelerometer sensor in any direction, the acceleration magnitude is loaded on the axes along which the motion is performed, plus the gravity value. The output of the accelerometer is a strip sequence of 3D points called G, such that a point over G is represented by ht{ax,ay,az}. The time stamp generated automatically by the sensor is denoted by ht, and records from the accelerometer at time t are denoted by $\{a_x, a_y, a_z\}$. The angle of tilting among the accelerometer axes is calculated as follows.

$$Angle = \tan^{-1}\left(\frac{\sqrt{(x^2 + y^2)}}{z}\right)$$

We extended ubiquitous system called the UbiGesture [5] that captures user context in ubiquitous environments. The context of the user is composed of location, user preferences, device, application and object for interaction. Figure 2 illustrates the system overview. UbiGesture will record the captured context parameters in shared storage, so it can be accessed by the proposed tilting interface. Once the interface receives the context environment parameters, it will load the appropriate tilt gestures for the user. The interface connects to the 3D accelerometer and begins tilt gesture recognition.

### 2.1. Pre-processing accelerometer data

There are two approaches for capturing gestures using accelerometers, discrete and continuous gesture recognition [23]. Discrete gestures must be initiated by the user, who determines the start and end of each gesture by hitting buttons or clicking. On the other hand, continuous gesture recognition is captured directly from user hand movements. Using the discrete gesture recognition is not intuitive for users as they must use another device or remain focused on precisely when to hit and release the button. Thus, we used the continuous gesture recognition to enrich the interaction with devices. After making the tilt gesture, however, the user must hold the position for a period (less than 200 ms) for the movement to be recognized by the system. This value is customized by the user depending on the interaction application.

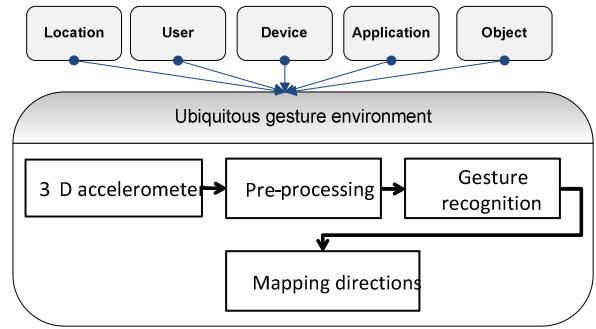

Figure 2. System Overview

To ensure that the system captures the correct tilt gesture performed by the user, the system keeps a continuous record of six successive accelerometer points. The pre-processor calculates the standard deviation of the six points; if the standard deviation is small, this means that the points are near one another. Thus, it indicates that the user's tilt hand is fixed in this direction, and the position is not classified as noise. The system calculates the average of the six points as the user's tilt gesture point in 3D space.

## 2.2. Gesture recognition

Because the initial state for the sensor is not defined right from the beginning of usage, the system must calibrate the initial steady state position. The system stores information for each application, object, available directions and calibrated points for each direction. This information is provided to the interface through the UbiGesture system. The tilt directions are calibrated as points in 3D space; thus, each point must satisfy two rules to be calibrated. First, the point must be fixed in the direction for a certain amount of time. Second, the minimum distance between the point and all other calibrated points must be larger than the virtual border value. The virtual border value is calculated as the maximum value of G applied over the x, y, z axes divided by desired calibrated directions. When the user performs a new tilting gesture, the system first ascertain whether the user's hand is intentionally fixed in this position, and then it will calculate the minimum distance with all the stored calibrated points, and returns the appropriate direction.

## 3. APPLICATION EXAMPLES

We studied some of the applications that could be suited for tilt gesture mapping. Applications as Google Earth, web browsers, media players, PTZ Camera Protector, special Japanese text tool called "Popie" [24], presentation viewers, image viewers and several gaming programs were tested with tilt gestures. We have selected three samples of application types that range in function and are suitable for remote interactions to evaluate the usability of our system.

## 3.1. Computer-based presentation viewer

Computer-based presentation viewers are commonly used by many users. Currently, a user must bend over to look for an icon on the display to change the slides, or press a button on the keyboard to control the slides. Other alternatives for interaction are using a remote mouse or asking an assistant to flip slides, both of which interrupt the flow of a presentation. In this type of situation, hand gestures could be appropriate, as they are a method of non-verbal communication between the presenter and audience. The presenter, however, cannot execute full hand gestures to flip slides, as this could be disruptive. Hence, tilt gestures are useful because they can be performed subtly in a small space. In our experiment, we assume that the user has created his presentation slides with commercial presentation software such as Microsoft's PowerPoint. The user attaches the sensor directly to his hand or another object and can browse through the presentation slides directly by performing tilt gestures.

Table 1.  Slideshow operations and mapping gestures.

| Operation | Description | Tilt gesture |
|-----------|-------------|--------------|
| Next | Advance to next slide | Right |
| Previous | Return to previous slide | Left |
| Home | Go to first slide | Up |
| End | Go to last slide | Down |
| Close | End slide show | Down-right |

We presented a method for users to interact with presentation slides through tilt gestures by executing two successive tilt motions in the desired direction, and then returning to the initial position. Figure 3 shows a user browsing presentation slides backwards and forwards. We choose one-to-one direct mapping between operation and gestures, so users easily manipulate the slides with a minimum number of gestures. The reverse-operation nature of browsing PowerPoint slides as next/previous slides is mapped directly to right/left tilt gestures. Table 1 shows the slide-show operations and mapping gestures. The "close" operation is mapped to the down-right tilt direction, as it was shown in the results of Andrew et al. [20] that the down-right tilt direction was the easiest diagonal direction for right handed users.

### 3.2. Photo browsers

Photo browsing applications are popular and are used in various situations. People like to browse through pictures with their families on large displays in living rooms. Students like to share pictures among each other using mobile devices. We implemented a photo browsing application that displays the stored pictures in rows and columns. Table 2 shows the mapping between operations and gestures.

Table 2.  Photo browser application operations and tilt gestures.

| Operation | Description | Tilt gesture |
|-----------|-------------|--------------|
| **Browsing mode** | | |
| Move right | Advance cursor 1 column | Right |
| Move left | Return cursor 1 column | Left |
| Move up | Advance cursor 1 row | Up |
| Move down | return cursor 1 row | Down |
| View picture | Open picture in new window | Down-right |
| **Editing mode** | | |
| Increase brightness | Increase brightness value | Up then right |
| Decrease brightness | Decrease brightness value | Up then left |
| Black and white filter | Greyscale the picture | Down |
| Close | Close current window | Left |

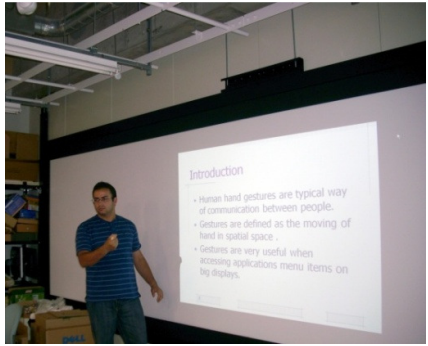 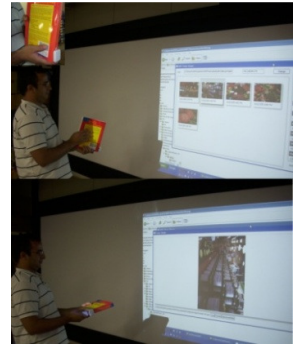

Figure 3. User browsing presentation slides    Figure 4. User selecting picture from grid, and opening the picture in a new window.

The system maps the operations to move right, left, up, and down as direct mapping for tilt gestures, right, left, up, and down respectively. There is also a function for adjusting the brightness of an image by a certain value. The user opens the image in a new window and selects the up tilt gesture to select the 'adjust image brightness' function. Then, he/she either increase or decrease the brightness value by right and left tilting gestures. Figure 4 shows a user selecting a picture from a grid and opening the picture in a new window. The combinations of tilt up then tilt right or left are triggers to execute commands.

### 3.3. Gaming systems

Gaming systems are an attractive field for many researchers to study user feedback and usability of systems. We tested an example of gaming interaction using a large screen display in our laboratory and a flight simulator game. Figure 5 shows a user interacting with flight simulator game.

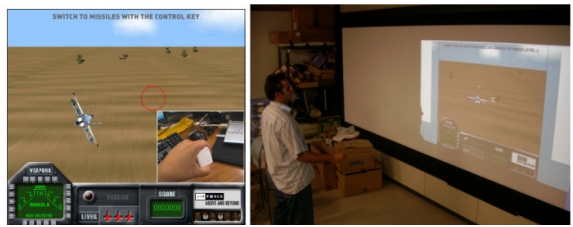

Figure 5. Left: tilting sensor to control flight speed; Right: user interacting with flight simulator.

The game is operated by four directional movements to control the mouse cursor on the screen. The user attaches the sensor to a flight object toy. The system extracts motion while interacting with the remote display screen. The displacement value is calculated according to the captured acceleration, screen resolution factor, and angle of the three axes of the accelerometer. The interface then calculates remote screen mouse display positions. The more the user tilts the object in one direction, the faster the flight speed on the screen becomes. The centre point of the remote display screen is the initial starting point for cursor movement. This displacement value is used to set the new position on the remote display screen. If the user holds the sensor horizontally towards the ground, the cursor will not move. Table 3 shows the flight simulator commands and mapped tilt gestures.

Table 3.  Flight simulator game operations and gestures.

| Operation | Description | Tilt gesture |
|-----------|-------------|--------------|
| Move right | Advance mouse cursor over x-axis | Right |
| Move left | Reduce mouse cursor over x-axis | Left |
| Move up | Reduce mouse cursor over y-axis | Up |
| Move down | Advance mouse cursor over y-axis | Down |

# 4. MAPPING DIRECTIONS

We have classified the mapping of tilt gestures according to application needs. Tilt gestures are mapped to application commands in means of direct, sequence and mouse mapped.

## 4.1. Direct mapping

A normal user can make hand tilt gestures in two, four, six or eight directions. A typical movement for direct mapping gesture is composed of two movements. First, the user must start from a steady state, and choose the direction of movement. Second, the user must return to the steady state. In the direct mapping, each tilt gesture is mapped to one command. If the user wants to move in more over than eight directions, he/she could lose sense of the direction of hands in space. The limits of the space in which to move is one of the main challenges of hand tilt.  To overcome the limited space of hand-tilt gestures, we presented the concept of the tilt-gesture level, so that each of the eight basic directions can have up to three levels of tilting. Figure 6 shows three levels for the up-right tilt gesture.

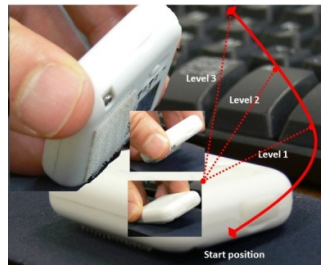

Figure 6. Tilt levels for the up-right direction.

This combination could allow us to provide up to 24 gestures for an application (eight directions with three levels each). Several commercial and developed applications can utilize this type of method, such as presentation viewers, photo viewers, media players and map-browsing applications like Google Earth. Also some applications like PTZ (pan, tilt, zoom) CameraProtector application requires the use of many functions, such as controlling PTZ or redirecting the camera to predesignated targets.

## 4.2. Tilt sequence

In some situations, when too many commands are used to control an application, a sequence of successive tilt gestures can be used. A tilt gesture sequence is built from two, three, four, or even more sequential tilt gestures. First, the user starts from a steady state and selects the first direction. Then, the user performs one or more other directions. The interface keeps a record of this sequence until the user has finished the sequence by returning to the steady state position. Figure 7 shows the process of building a tilt sequence for an up then right sequence. The interface will parse the sequence and trigger the functions. This mode is suitable for applications with many functions that depend on adjusting values or executing a sequence of commands. An example of applications that could use tilt sequence is editing photos and applying various filters. Furthermore, special text editors that depend on a sequence or pattern of movements are adequate for this tilt method.

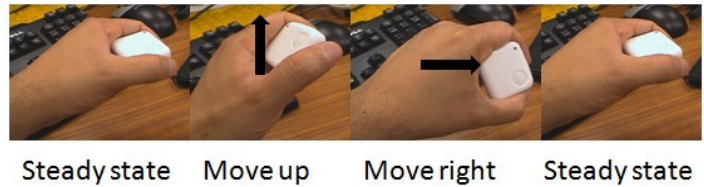

Steady state    Move up        Move right      Steady state

Figure 7. Tilt sequence, building a sequence of tilt movements up, followed by right.

## 4.3. Mouse movement

Some applications require movement of the mouse to control a cursor or objects on the screen. In this case direct mapping or tilt sequences are not sufficient because they are designed for triggering events to execute commands but cannot directly control mouse movements across the screen. Tilt gestures can have some feedback thought the control of the 2D (x, y) mouse coordinates across the display. There exist two methods to extract mouse motion from 3D accelerometer sensors. The first method is by calculating a displacement value relative to the angle of tilt among the 3D accelerometer axes. The second method is by depending on threshold values over the x and y axes, and then calculating a displacement speed from static acceleration. In both methods, screen resolution is a factor for smooth movement of mouse curser over the display. To perform the mouse left clicks, right clicks, and scrolls using continuous gesture recognition, we use a type of 'tap' or 'shake' operation. This technique is best suited for applications that depend on mouse movements, like gaming programs such as flight simulators, car-racing games, and drawing applications.

## 5. EVALUATION

To evaluate the usability of the tilting interface while interacting with graphical user interface items (GUI), we built a circle menu in which each piece virtually represents application functions or GUI items, such as buttons, icons and thumbnails. The circular menu can display up to 16 circular sequences of targets. Figure 8 displays the experimental circular menu in different sizes. The target item is shown on the menu by the highlighted light-blue colour. Every time the subject hits a target successfully a new target appears in a different place. If the subject hits unmarked target items, this is counted as an error. Time is measured as the time in seconds needed to move from the steady state position to hit the current target.

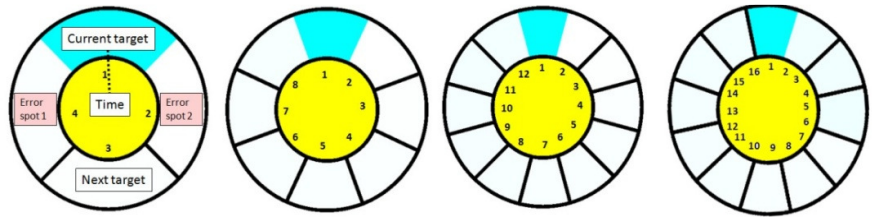

Figure 8. Developed experimental application with different target numbers and sizes.

Twelve subject aged 22 to 30 participated in the experiment. Subjects were divided into two groups to test the learning effect. Each group was assigned to do the task in a given sequence.

First group was asked to interact with a number of targets in an ascending order {4, 8, 12 and 16} targets. Second group was asked to interact in descending order {16, 12, 8, and 4}. At the end of the experiments, we make a short interview with the subjects to get their feedback, and we asked them to fill a satisfaction survey. We divided the 12 subjects into more 3 groups for using tilting interface with different objects. The first group did not use any object for holding the sensor, just using hands, the second group used ball and third group used a stick. We selected those objects because of their individual geometrical shape and hand holding control. Figure 9 shows objects used in our experiment.

The experiment concerns the measurement of accuracy, speed, learning of directions to practice and the appropriate number of tilt directions to be recognized. Each subject has to perform two sessions per each interaction method. Each session composed of {4,8,12,16} targets selections, a total of 40 targets per session. At the beginning of each session, we ask the subjects to make a practice session for hitting the targets using the three methods.

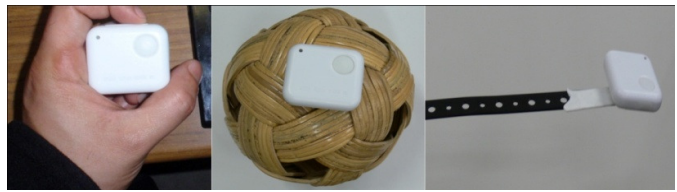

Figure 9. 3D accelerometer sensor attached to hands, ball and stick objects.

## 5.1. Results

The results indicate that the use of conventional interaction methods such as mouse is the fastest methods for interacting with the experimental application. However, the results founded that tilt gestures have a stable speed regardless the number of targets. The hand gestures were the lowest in speed and show a linear increase of speed with respect to the number of targets. Figure 10 (a) shows the speed comparison between mouse, hand gesture and tilt interface. The accuracy to hit targets while using the tilt interface was close to mouse for the {4, 8} targets, but the results show a linear decrease of accuracy when the number of targets increased to {12, 16}. Figure 10 (b) shows the accuracy comparison between mouse, hand gestures and tilt interface. The accuracy for hand gestures was less than tilt and mouse, but the results show that for the number of targets of {8, 12, 16} the accuracy was stable on average of 75%. However, tilt gesture accuracy for 16 targets was still more accurate than hand gestures.

The tilt interface took more time than that of the mouse because of subjects sometimes either forgets to return to steady state after hitting the target or they miss the position of steady state. The declination of accuracy has two reasons, first was the difficulty to imagine the direction levels for each direction without having a direct visual feedback from the interface. Second, subjects find it hard to make 12 gestures or more using wrist movement only and direct mapping. The hand gestures consumed a lot of time to hit the targets and accuracy was the minimum. The reason behind this is because first, the subjects took time to understand the shape of each gesture, especially gestures done by subjects towards the down directions. Second, the subjects felt tired and complained from hand tremor while using hand gestures, leading them to make wrong hand gesture patterns.

We have done further analysis for the speed comparison between the devices with respect to learning effect for Group1 {4, 8, 12, 16} and Group2 {16, 12, 8, 4}. The results founded that there is a difference of 1 second between groups for the hand gestures. However, the learning effect for tilt gestures remains constant for both Group1 and Group2. Stability in learning effect is an advantage of tilt gestures over hand gestures as subjects can easily memorise direction of tilts. When subjects used the hand gestures they consume less learning time when they start from the large amount of targets (16) towards (4) targets as they learn most of the hand gestures in the session of 16 targets.

We studied the effect of objects on the tilting interface for subjects while they interact with the experimental menu. We ask the subjects to fill in a satisfaction survey for each method and objects used and give a rate on a scale from 1 to 5. It was observed that handling thin objects using the thumb and index finger only was not appropriate for tilt interaction. They allow the subjects to move the object in all the spatial space around their wrist leads to misinterpretation of tilting directions.

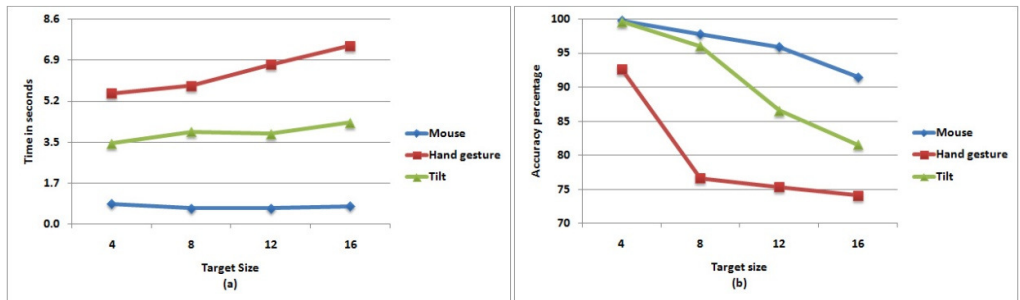

Figure 10. Comparison between mouse, hand gestures and tilt interface (a) speed (b) accuracy.

We analyzed the satisfaction between objects, and we found that subjects were less satisfied for stick object. The ball and hand got the same satisfaction percentage for the 12 targets around 65%. Subjects say that they feel more comfortable while holding the ball and they feel easy to control. The stick object satisfaction was far away from what we expected at 16 targets, as it was difficult to perform levels using thin objects with large spatial space around. Figure 11 shows satisfaction percentage for each object and the number of targets.

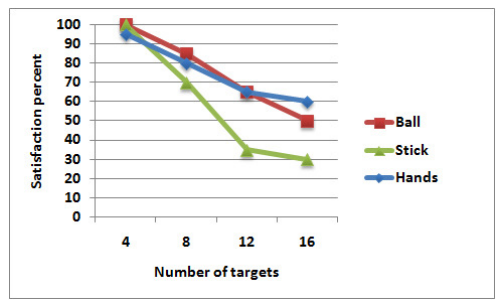

Figure 11. Subjects satisfaction percent per number of targets and objects.

The subjects have given feedback during an interview about the use of tilting interface after finishing the experiments. Subjects make the move left full hand gesture as tilt to left. Subjects indicate that using the tilting interface was easy and natural to use and less painful in comparison to full hand gestures. Subjects cannot imagine how to tilt the sensor in 3D-space and the mapping towards the experimental application in 2D directions. All subjects agree that the mouse was the most comfortable tool for interaction, but it lacks remote interaction and should have a stable hand position which is unsuitable for on the move interaction. Some subjects confirm that using objects around as balls, pens, and headphones could enrich their interaction with devices around in the ubiquitous computing environment.

# 6. CONCLUSIONS

Tilting gestures were studied as an interaction method in ubiquitous environments where it is difficult to have ordinary input devices. Tilt gestures were combined with objects to control applications and devices. In this paper, we proposed a system that captures people context and provides them tilt gestures for objects they interact with. We discuss the system settings, gesture recognition and methods for mapping directions using a 3D accelerometer to capture the tilting gestures.

We explored application fields and areas were tilting gestures could be useful by applying it to map and web browsers, media players, PTZ camera protector and special Japanese text tool. The usefulness and feedbacks of the tilt gesture interface were evaluated by studding it with three different types of remote interaction applications (computer presentation viewer, photo browsers, and entertainment games).

We present the mapping techniques for each application commands and tilt gestures based on objects they use for interaction and some user's feedback. We addressed the classification of suitable applications for tilting interfaces as direct mapping, tilt sequence and mouse moves. Furthermore, we need to study other types of tilt gestures mapping techniques based on detailed long term usage of the system.

An experiment to evaluate the interaction with a menu representing GUI items was conducted. The results show that tilt interface was close to mouse in accuracy and speed for up to 8 targets. Even when the number of targets increased up to 16 targets tilt gestures still perform much better than hand gestures. Also we found that tilt gestures were easy to memorize by most of the users, whether they start by 4 or 16 targets there was stability in results. Furthermore, a group of users feedback showed an agreement on using objects around them enrich their interaction with applications like using balls, pens and headphones. However, we want to evaluate the usage of tilt levels as a method to increase the number of commands to be executed by tilt gestures.

**Authors**

Ayman Atia    received the BSc and MSc. degree from the Department of Computer Science, Helwan University, Egypt in 2000 and 2004, respectively. He is currently pursuing his Ph.D. at the Department of Computer Science, University of Tsukuba, Japan. His research interests include human computer interaction, wireless sensors, and interactions in ubiquitous environments with everyday objects. He is a student member of the IEEE computer society and IPSJ.

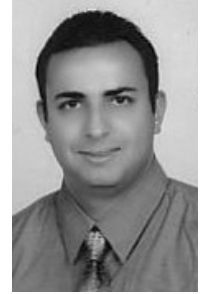

Jiro Tanaka    is a professor in the Department of computer science, Graduate school of systems and information engineering, University of Tsukuba. His research interests include visual programming, interactive programming, computer-human interaction and software engineering. He received a BSc and MSc from the university of Tokyo in 1975 and 1977. He received a PhD in computer science from the university of Utah in 1984. He is a member of the ACM, the IEEE computer society and IPSJ

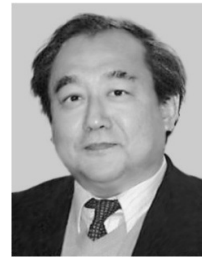